\documentclass[twocolumn,showpacs,preprintnumbers,amsmath,amssymb]{revtex4}
\usepackage{graphicx}
\begin{document}
\title{Quantum walks as a probe of structural anomalies in graphs}
\author{Mark Hillery and Hongjun Zheng}
\affiliation{Department of Physics, Hunter College of the City University of New York, 695 Park Avenue, New York, NY 10065 USA}
\author{Edgar Feldman}
\affiliation{Department of Mathematics, Graduate Center of the City University of New York, 365
Fifth Avenue, New York, NY 10016 USA}
\author{Daniel Reitzner} 
\affiliation{Department of Mathematics, Technische Universit\"at M\"unchen, 85748 Garching, Germany}
\author{Vladimir Bu\v{z}ek}
\affiliation{Research Center for Quantum Information, Slovak Academy of Sciences, D\'ubravsk\'a cesta 9, 845 11 Bratislava, Slovakia}

\begin{abstract}
We study how quantum walks can be used to find structural anomalies in graphs via several examples.  Two of our examples are based on star graphs, graphs with a single central vertex to which the other vertices, which we call external vertices, are connected by edges.  In the basic star graph, these are the only edges.  If we now connect a subset of the external vertices to form a complete subgraph, a quantum walk can be used to find these vertices with a quantum speedup.  Thus, under some circumstances, a quantum walk can be used to locate where the connectivity of a network changes.   We also look at the case of two stars connected at one of their external vertices.  A quantum walk can find the vertex shared by both graphs, again with a quantum speedup.  This provides an example of using a quantum walk in order to find where two networks are connected.  Finally, we use a quantum walk on a complete bipartite graph to find an extra edge that destroys the bipartite nature of the graph.  
\end{abstract}


\maketitle

\section{Introduction}
One of the most versatile quantum algorithms is the quantum search algorithm due to Lov Grover \cite{grover}.  In
its original form, it identified which Boolean function from a particular set was realized by a particular quantum
oracle.  A Boolean function, $f(x)$ maps $n$ bit binary numbers to either $0$ or $1$, and the particular class of 
Boolean functions considered by the simplest form of the Grover algorithm are $0$ for all strings except one.  We
are given an oracle that realizes one of these functions; if we input $x$ its output is $f(x)$.  Our task is to find
which function it realizes, or, equivalently, for which input $f(x)=1$, with as few calls to the oracle as possible.  
Classically one needs of order $2^n$ calls whereas on a quantum computer, using the Grover algorithm, one
needs only of order $2^{n/2}$ calls.

A variant of the Grover algorithm was defined for searches on graphs.  First, one defines a quantum walk on a
graph, which is a quantum version of a random walk \cite{walk1}-\cite{walk5}.  Then the behavior of one of the
vertices is changed so that it acts differently from all of the others.  The object is then to find the distinguished
vertex.  This has been done for a number of highly symmetric graphs, such as the hypercube \cite{shenvi,potocek}, grids in different dimensions \cite{grid1,grid2}, and the complete graph \cite{grid2,complete2}.  The initial state of the walk cannot incorporate any knowledge of the distinguished vertex, and it is usually an equal superposition of all vertices, in the case of a coined walk, or an equal superposition of all edges, in the case of a scattering walk.  The number of steps the walk must take in order to find the distinguished vertex is of the order of the square root of the number of vertices in the graph. Some of the latest studies of searches on graphs have focussed on how the search is affected by the connectivity of the graph or by disorder in the graph \cite{kendon2}, or searching in a graph in which there are several kinds of non-special vertices \cite{lee}.  It should be noted that by constructing a 
quantum circuit that implements a quantum walk, these graph search problems can be rephrased as searches
involving calls to an oracle.  For an explicit example of this see \cite{complete2}.

More recently, it has been found that quantum walks can find things besides distinguished vertices in a graph
\cite{feldman}.  In that study, walks on star graphs were examined.  A star graph has a central vertex and $N$
edges emanating from it, each of which is connected to its own vertex, so that the graph has a total of $N+1$
vertices.  We shall call the vertices besides the central vertex external vertices.  If one adds an extra edge 
connecting two of the external vertices, it is possible to find the extra edge in approximately $\sqrt{N}$ steps
using a quantum walk.  If one adds a loop to an external vertex, the result is the same, but if one adds a new
vertex and an edge between that vertex and one of the external vertices, the quantum walk does not find the
extra edge.  So, it is unclear what kinds of structural anomalies can be found and what kinds cannot.

Here we would like to continue our exploration of this subject.  We begin by reviewing some of the results of
\cite{feldman} and presenting more details.  We then move on to several more examples.  First we consider
a star graph with extra edges added connecting external vertices so that these external vertices form a complete
graph.  A complete subgraph of a graph is known as a clique.  The idea is to use a quantum walk to find the
vertices that comprise the clique.  Note that what this does is allow us to find a part of the graph in which the structure of the network changes.  In the star graph, the external vertices are only connected to each other through the central vertex.  If we now form a region in which the density of connections increases, in particular in which the external vertices are directly connected to each other, we can use a quantum walk to find this region.  Next, we consider two star graphs joined at one of their external vertices.  In this case the walk starts on both of the stars, and we want to find the vertex where the two star graphs are connected.  This shows that we can use a quantum walk to find where two networks are connected to each other. Finally, we look at the case of a complete bipartite graph.  In this graph, the vertices are divided into two sets, and each vertex in one set is connected to all of the vertices in the other set by an edge, but no vertices within the sets are connected.  Suppose we now add one edge connecting two vertices in one of the sets.  We can use a quantum walk to find this edge faster than we could classically.  

\section{Star graphs}
Throughout this paper we will be using the scattering quantum walk in which the particle ``scatters'' off the vertices of the graph \cite{walk4,complete2}.  There is another version of the discrete-time quantum walk, known as a coined walk \cite{walk1,walk2}.  In this type of walk, there is an extra system, the coin, that makes the step transformation unitary and controls the dynamics of the walk.  The coined walk has been shown to be equivalent to the scattering walk, so that which one is used is a matter of preference \cite{andrade}.  We find the scattering walk more physically motivated, and it is the one we shall use.  In this walk the particle making the walk sits on the edges of the graph instead of the vertices.  Each edge has two orthogonal states.  If the edge connects vertices $j$ and $k$, then one state is $|j,k\rangle$ corresponding to the particle going from $j$ to $k$, and the other is $|k,j\rangle$ corresponding to the particle going from $k$ to $j$.  The collection of all of these states, two for each edge, forms an orthonormal basis for the Hilbert space in which the walk takes place.  In addition to the Hilbert space we need a unitary operator that advances the walk one step.  In the scattering walk each vertex acts as a scattering center and is described by a local unitary operator that maps states entering the vertex to states leaving the vertex.  The unitary operator that advances the walk one step, $U$, is simply made up of the action of all of the local unitary operators at the vertices. For a vertex, $j$, with $n$ edges connected to it, we will generally use the operator
\begin{equation}
U|k,j\rangle = -r |j,k\rangle + t \sum_{l=1,l\neq k}^{n} |j,l\rangle ,
\end{equation}
where $r=(n-2)/n$ and $t=2/n$.  This type of vertex behaves in the same way no matter from which edge it is entered.  

As was stated in the Introduction, a star graph has a central vertex, which we shall denote by $0$, and $N$ external vertices, which we shall denote by $1$ through $N$.  Each of the external vertices is connected to the central vertex by a single edge.  The dimension of the Hilbert space in which a walk on this graph takes place is $2N$.  The central vertex behaves as described in the preceding paragraph, and the behavior of the external vertices depends on the application.  Since the walk on a star graph with an extra edge was discussed thoroughly in \cite{feldman}, here we will describe what happens when we add loops to the external vertices.

Let us first consider the case in which we add a loop to a single external vertex, which we shall take to be vertex $1$.  We shall denote the single state of the loop by $|l_{1}\rangle$.  The operator $U$ now acts on the states entering the external vertices as $U|0,1\rangle = |l_{1}\rangle$, $U|l_{1}\rangle = |1,0\rangle$, and $U|0,j\rangle = |j,0\rangle$ for $j\geq 2$.  Let us now define the states
\begin{eqnarray}
|\psi_{1}\rangle & = & \frac{1}{\sqrt{N-1}}\sum_{j=2}^{N} |0,j\rangle  \nonumber \\
|\psi_{2}\rangle & = & \frac{1}{\sqrt{N-1}}\sum_{j=2}^{N} |j,0\rangle  ,
\end{eqnarray}
and note that 
\begin{eqnarray}
U|1,0\rangle & = & -r|0,1\rangle + t\sqrt{N-1}|\psi_{1}\rangle \nonumber \\
U|\psi_{1}\rangle & = & |\psi_{2}\rangle \nonumber \\
U|\psi_{2}\rangle & = & t\sqrt{N-1}|0,1\rangle +r |\psi_{1}\rangle .
\end{eqnarray}
Now,  if $S$ is the subspace spanned by the vectors $\{ |0,1\rangle , |l_{1}\rangle , |1,0\rangle , |\psi_{1}\rangle , |\psi_{2}\rangle \}$, we note that $S$ is invariant under the action of $U$.  This implies that if the initial state of the walk is in $S$, the entire walk will take place in $S$, which reduces the dimension of the space we have to consider from $2N$ to $5$.  This will be a feature of all of the problems we consider here, a drastic reduction in the size of the space due to the high symmetry of the graph.  This type of dimensional reduction was
first used by Krovi and Brun in studies of coined quantum walks \cite{krovi}.  In this case, it means that $U$ restricted to $S$, $U_{S}$, is given by the $5\times 5$matrix
\begin{equation}
U_{S} = \left( \begin{array}{ccccc} 0 & 0 & -r & 0 & t\sqrt{N-1} \\ 1 & 0 & 0 & 0 & 0 \\ 0 & 1& 0 & 0 & 0 \\ 0 & 0 & 
t\sqrt{N-1} & 0 & r \\ 0 & 0 & 0 & 1 & 0 \end{array} \right) ,
\end{equation}
where the basis is ordered as above in the definition of $S$.  The state of the walk after $n$ steps will be $U_{S}^{n}|\psi_{init}\rangle$, where $|\psi_{init}\rangle$ is the initial state of the walk and we shall assume that $|\psi_{init}\rangle \in S$.  In order to evaluate this, we want to find the eigenvalues and eigenstates of $U_{S}$.  

The characteristic polynomial of $U_{S}$ is 
\begin{equation}
\lambda^{5} -r\lambda^{3} + r\lambda^{2}- 1= (\lambda -1)(\lambda^{4}+\lambda^{3}+t\lambda^{2} + \lambda +1)
=0 .
\end{equation}
We see immediately that $\lambda = 1$ is a root, but in order to find the others, we shall resort to perturbation theory.  If $N\gg 1$, then $t\ll 1$, and to find our zeroth order solution we set $t$ equal to $0$.  The equation for the remaining roots then becomes $(\lambda +1)(\lambda^{3}+1)=0$, so that, to zeroth order, the remaining roots are $-1$ (twice) and $\exp (\pm i\pi /3)$.  Now we need to find the lowest order corrections to these eigenvalues.  It turns out that the only interesting eigenvalue is $-1$.  This is because, as we shall see, the corrections to $-1$ are $O(N^{-1/2})$ while the corrections to the other eigenvalues are $O(N^{-1})$.  In order to obtain a quantum speedup, we need the state to change substantially in $O(N^{1/2})$ steps.  This will happen for a superposition of eigenstates whose eigenvalues are of the form $\lambda_{0} + O(N^{-1/2})$, where $\lambda_{0}$ is the zeroth order eigenvalue, but not for superpositions of states whose eigenvalues are of the form $\lambda_{0} + O(N^{-1})$.  In order to find the lowest order corrections to $-1$ we set $\lambda = -1 + \delta\lambda$ and substitute it back into the fourth order equation for $\lambda$ keeping only lowest order term.  We find
\begin{equation}
\delta\lambda = \pm i\sqrt{\frac{t}{3}} ,
\end{equation}
which is $O(N^{-1/2})$.  The corresponding eigenvectors are
\begin{equation}
|v_{+}\rangle = \frac{1}{\sqrt{6}} \left(\begin{array}{c} 1 \\ -1 \\ 1 \\ -i\sqrt{3/2} \\ i\sqrt{3/2} \end{array}\right) 
\hspace{5mm}
|v_{-}\rangle = \frac{1}{\sqrt{6}} \left(\begin{array}{c} 1 \\ -1 \\ 1 \\ i\sqrt{3/2} \\ -i\sqrt{3/2} \end{array}\right) ,
\end{equation}
with $|v_{+}\rangle$ corresponding to $-1+i\sqrt{t/3}$ and $|v_{-}\rangle$ corresponding to $-1-i\sqrt{t/3}$.

Now, for the initial state of our walk, let us choose the state
\begin{eqnarray}
|\psi_{init}\rangle & = & \frac{1}{\sqrt{2N}} \sum_{j=1}^{N}(|0,j\rangle - |j,0\rangle ) \nonumber \\
& = & \frac{1}{\sqrt{2N}}(|0,1\rangle - |1,0\rangle )  \nonumber \\
& & + \sqrt{\frac{N-1}{2N}}(|\psi_{1}\rangle - |\psi_{2}\rangle )  ,
\end{eqnarray}
which we can see is in $S$.  Noting that
\begin{equation}
|\psi_{1}\rangle - |\psi_{2}\rangle = i(|v_{+}\rangle - |v_{-}\rangle ) ,
\end{equation}
we see that that the initial state is approximately equal to a superposition of two eigenvectors
\begin{equation}
|\psi_{init}\rangle = \frac{i}{\sqrt{2}} (|v_{+}\rangle - |v_{-}\rangle ) + O(N^{-1/2}) .
\end{equation}
If we now express $-1\pm i\sqrt{t/3}\cong - \exp (\mp i\theta)$, where $\theta = \sqrt{t/3}$ we find that
\begin{eqnarray}
U^{n}|\psi_{init}\rangle & \cong & \frac{i}{\sqrt{2}} (-1)^{n} (e^{-in\theta}|v_{+}\rangle - e^{in\theta} |v_{-}\rangle )
\nonumber \\
& \cong & \frac{(-1)^{n}}{\sqrt{3}} \left( \begin{array}{c} \sin (n\theta ) \\ -\sin (n\theta ) \\ \sin (n\theta ) \\
\sqrt{3/2}\cos (n\theta ) \\ -\sqrt{3/2}\cos (n\theta ) \end{array}\right) .
\end{eqnarray}

Examining the form of $U^{n}|\psi_{init}\rangle$, we see that when $n\theta = \pi /2$ (this implies $n$ is $O(N^{1/2})$), the particle is either on the edge connected to the loop, with probability $2/3$, or on the loop itself, with probability $1/3$.  Now, in measuring where the particle is, we assume that we do not have access to the loop, otherwise we would know where it is, but we do have access to all of the edges.  Therefore, if we measure the position of the particle after a number of steps satisfying $n\theta = \pi /2$, we will with a probability of $2/3$ find the particle on the edge connected to the loop.  With a probability of $1/3$, however, we will find no particle at all, and in that case we run the walk one more step, after which the particle will be on the edge connected to the loop.  Therefore, by running the walk for $O(N^{1/2})$ steps, we have found which edge is connected to the loop with a probability close to $1$.

In comparing this procedure to a classical search for the loop, we shall assume that classically
the graph is specified by an adjacency list, which is an efficient specification for sparse graphs.
For each vertex of the graph, one lists the vertices that are connected to it by an edge.  This list
can include the vertex itself, which means that there is a loop connected to that vertex.   Searching 
this list classically would require $O(N)$ steps to find the loop, while the quantum procedure will 
succeed in $O(\sqrt{N})$.

The pattern of this calculation will be repeated for the other examples we discuss.  First one finds an invariant subspace of small dimension in which the walk takes place.  Next, one diagonalizes the unitary operator that advances the walk one step, $U$, in that subspace.  This typically involves a perturbative approach to finding the eigenvalues and eigenstates.  The zeroth order solutions are found by looking at the $N\rightarrow \infty$ limit, and the small parameter in which one does the perturbation expansion is a power of $1/N$.  It is the eigenvalues that are degenerate to zeroth order that lead to the interesting parts of the Hilbert space.  One then identifies an appropriate initial state, and calculates the action of $U^n$ on that state.  Since this pattern holds for all of our calculations, we will present mainly the results in the body of the paper, and describe some of the details in the Appendix. 

Before leaving the star graph proper, let us look at one more example, which was also discussed in \cite{feldman}.  Suppose that all of the external vertices except one, which we shall take to be vertex $1$, have loops, and we want to find which vertex does not have a loop.  Actually, we have to be a bit more careful in our description, because we are now going to assume we have access to the loops, so if there were one missing, we would know where it is.  What we assume is that all of the external vertices are connected to loops, but the one connected to vertex $1$ is a dummy loop.  In particular, we assume that $U|0,j\rangle = |l_{j}\rangle$ and $U|l_{j}\rangle = |j,0\rangle$ for $j\geq 2$, and for vertex $1$, $U|0,1\rangle = e^{i\phi}|1,0\rangle$ and $U|l_{1}\rangle = |l_{1}\rangle$.  One only gets a quantum speedup for particular values of $\phi$.

This also reduces to a five-dimensional problem.  The invariant subspace in this case is spanned by the vectors $\{ |0,1\rangle , |1,0\rangle , |\psi_{L}\rangle , |\psi_{1}\rangle , |\psi_{2}\rangle \}$, where $|\psi_{1}\rangle$ and $|\psi_{2}\rangle$ are as before, and
\begin{equation}
|\psi_{L}\rangle = \frac{1}{\sqrt{N-1}} \sum_{j=2}^{N} |l_{j}\rangle .
\end{equation}
We find that the characteristic equation of $U$ restricted to the invariant subspace only has double roots in the $N\rightarrow \infty$ limit if $\phi$ is $\pi$, $\pi /3$ or $-\pi /3$, and these are the values of $\phi$ for which we obtain a quantum speedup.  In the case that $\phi = \pi$ the appropriate initial state is given by 
\begin{equation}
|\psi_{init}\rangle  =  \frac{1}{\sqrt{3N}} \sum_{j=0}^{N}(|0,j\rangle + |j,0\rangle + |l_{j}\rangle ) ,
\end{equation}
and the particle becomes localized on the edge with the dummy loop after $n=(\pi /2)\sqrt{3/t} =O(N^{1/2})$ steps.  For $\phi = \pm \pi /3$ different initial states are required, but the results are qualitatively the same.  The details of the case $\phi = \pi$ are given in the Appendix.

\section{Star graph with a clique}
Now suppose that we start with a star graph with $N$ edges, and we add extra edges to it.  The case of one extra edge was dealt with in \cite{feldman}, but now we wish to add enough edges so that a subset of the external vertices form a complete graph, or clique (see Fig.~1).  In particular, we shall assume that vertices $1$ through $M$ form the clique, i.e.\ each of these vertices is connected to all of the other vertices in the set $\{ 1,2, \ldots M\}$ as well as to the central vertex.  We will also assume that $M\ll N$.  This graph can be viewed as a network in which most of the participants are only connected through the central vertex, but there is a subset of participants who are directly connected to each other.  The object is to find the vertices in the clique.

\begin{figure}
\includegraphics[scale=.5]{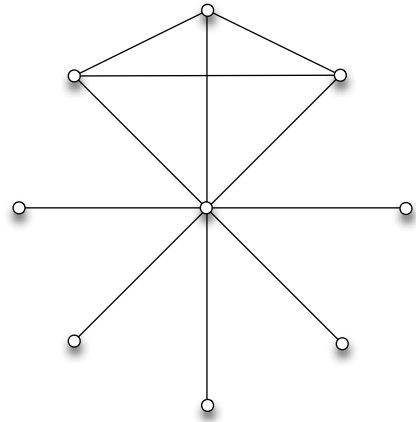}
\caption{A star graph with a clique, in this case a triangle.}
\label{star-clique}
\end{figure}

In this case, the operator $U$ acts as follows
\begin{eqnarray}
U|j,0\rangle & = & -r|0,j\rangle + t \sum_{k=1,k\neq j}^{N} |0,k\rangle \nonumber \\
U|0,j\rangle & = & -\tilde{r}|j,0\rangle + \tilde{t} \sum_{k=1,k\neq j}^{M} |j,k\rangle \hspace{3mm} 
{\rm for}\ 1\leq j \leq M \nonumber \\
U|0,j\rangle & = & |j,0\rangle \hspace{3mm} {\rm for}\ M+1\leq j \leq N \nonumber \\
U|j,k\rangle & = & -\tilde{r} |k,j\rangle +\tilde{t} |k,0\rangle + \tilde{t} \sum_{l=1, l\neq j, l\neq k}^{M} |k,l\rangle 
\nonumber \\
& & \hspace{3mm} {\rm for}\ 1\leq j,k \leq M  ,
\end{eqnarray}
where $r$ and $t$ are as before, and $\tilde{r}=(M-2)/M$ and $\tilde{t}=2/M$.  We choose the initial state to be
\begin{equation}
|\psi_{init}\rangle  =  \frac{1}{\sqrt{2N}} \sum_{j=1}^{N} (|0,j\rangle - |j,0\rangle ) ,
\end{equation}
and after running the walk for 
\begin{equation} 
n=\frac{\pi\sqrt{N}}{2} \sqrt{\frac{2M-1}{2M(M-1)}}  ,
\end{equation}
steps, the particle is located on one of the edges connecting the clique and the central vertex with a probability of $(2M-2)/(2M-1)$ (up to terms of order $(M/N)^{1/2}$) and it is located on one of the edges of the clique itself with a probability of $1/(2M-1)$.  We assume that we do not have access to the edges of the clique itself, so that when we measure the position of the particle we either find it on one of the edges emanating from the central vertex, or we don't find it at all, because it is on one of the edges of the clique.  Note that the probability of the particle being on one of the edges of the clique decreases as the size of the clique increases. 

Classically one would have to search the adjacency list of the graph in order to find a vertex that is a member of the clique, and one would have to check $O(M/N)$ elements.  This compares to the $O(\sqrt{M/N})$ steps the quantum walk must make in order to find one of the vertices in the clique.  Once one finds one vertex in the clique, the rest are found by reading off the vertices adjacent to that vertex from the adjacency list in both the classical and quantum cases.

\section{Two stars}
Now let us look at a different problem.  We have two stars, each with $N$ edges.  They share one external vertex, so the stars are connected, but we do not know which one (see Fig.~2).  The object is to find the shared vertex.

\begin{figure}
\includegraphics[scale=.5]{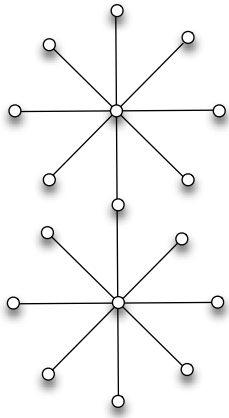}
\caption{Two star graphs connected at one of their external vertices.}
\label{two-star}
\end{figure}

Let us denote the central vertices of the two stars by $A$ and $B$.  In order to analyze a quantum walk on this graph, we shall assume that they share vertex $1$  The external vertices of the first star (with central vertex $A$) are $\{ 1,2,\ldots N\}$ and those of the second star (with central vertex $B$) are $\{ 1,N+1, N+2, \ldots 2N-1\}$.  The quantum walk in which we are interested takes place in an eight-dimensional invariant subspace spanned by the vectors $|\psi_{1}\rangle = |A,1\rangle$, $|\psi_{2}\rangle = |1,A\rangle$, $|\psi_{3}\rangle = |B,1\rangle$, $|\psi_{4}\rangle = |1,B\rangle$ and
\begin{eqnarray}
|\psi_{5}\rangle & = & \frac{1}{\sqrt{N-1}}\sum_{j=2}^{N} |A,j\rangle \nonumber \\
|\psi_{6}\rangle & = & \frac{1}{\sqrt{N-1}}\sum_{j=2}^{N} |j,A\rangle \nonumber \\
|\psi_{7}\rangle & = & \frac{1}{\sqrt{N-1}}\sum_{j=N+1}^{2N-1} |B,j\rangle \nonumber \\
|\psi_{8}\rangle & = & \frac{1}{\sqrt{N-1}}\sum_{j=N+1}^{2N-1} |j,B\rangle  .
\end{eqnarray}
These eight vectors form an orthonormal basis for the invariant subspace.  All of the vertices behave as before except for vertex $1$.  We shall assume that the particle is transmitted at vertex $1$, that is, there is no amplitude for it to be reflected there.  That means that
\begin{equation}
U|\psi_{1}\rangle = |\psi_{4}\rangle \hspace{5mm} U|\psi_{3}\rangle = |\psi_{2}\rangle .
\end{equation}
The operator that advances the walk one step acts on the other basis vectors in the invariant subspace as
\begin{eqnarray}
U|\psi_{2}\rangle & = &-r|\psi_{1}\rangle + t\sqrt{N-1}|\psi_{5}\rangle \nonumber \\
U|\psi_{4}\rangle & = & -r|\psi_{3}\rangle + t\sqrt{N-1}|\psi_{7}\rangle \nonumber \\
U|\psi_{5}\rangle & = & |\psi_{6}\rangle \nonumber \\
U|\psi_{6}\rangle & = & r|\psi_{5}\rangle + t\sqrt{N-1}|\psi_{1}\rangle \nonumber \\
U|\psi_{7}\rangle & = & |\psi_{8}\rangle \nonumber \\
U|\psi_{8}\rangle & = & r|\psi_{7}\rangle + t\sqrt{N-1}|\psi_{3}\rangle ,
\end{eqnarray} 
where, as before, $t=2/N$ and $r=(N-2)/N$.

We now start the particle in the state
\begin{eqnarray}
|\psi_{init}\rangle & = & \frac{1}{2\sqrt{N}} \left[ \sum_{j=1}^{N} ( |A,j\rangle + |j,A\rangle ) - (|1,B\rangle 
+ |B,1\rangle ) \right.  \nonumber \\
& & \left. - \sum_{j=N+1}^{2N-1}(|B,j\rangle + |j,B\rangle ) \right]  ,
\end{eqnarray}
that is, a superposition of all of the edge states in the first star minus a superposition of all of the edge states in the second.  We find that after $n=\pi\sqrt{N}/2$ steps, the particle is located with high probability ($1-O(N^{-1/2})$) on either the edge between vertices $1$ and $A$ or the edge between vertices $1$ and $B$.  Therefore, using the quantum walk we can find the external vertex the star graphs have in common with $O(\sqrt{N})$ steps, whereas classically we would have to search the adjacency lists of the external vertices of one of the stars, which means searching a combined list containing $N+1$ items (one item from each of the vertices connected only to the central vertex and two items from the vertex connected to both central vertices).  Therefore, the quantum walk gives us a quadratic speedup.  

\section{Complete bipartite graph}
We will now consider a type of graph that is actually a generalization of a star graph.  A bipartite graph is one in which the vertices are divided into two sets, and only vertices in different sets are connected by edges; there are no edges between vertices in the same set.  A complete bipartite graph is one in which each element in one set is connected to all of the elements in the other set.  In the case of a star graph, one set contains only the central vertex and the other contains the external vertices.  We shall assume that there are $N_{1}$ vertices in set $1$ and $N_{2}$ vertices in set $2$, so that there are $N_{1}N_{2}$ edges in total.  The vertices in set $1$ will be labelled $1,2,\ldots N_{1}$, and those in set $2$ will be labelled $N_{1}+1, N_{1}+2, \ldots N_{1}+N_{2}$.  Finally, we will add one more edge, between vertices $1$ and $2$.  This edge destroys the bipartite character of the graph. What we want to determine is whether a quantum walk can help us find this edge.  It could, in principle, be between any two vertices in set $1$ or between any two vertices in set $2$.  However, we are going to analyze a situation in which the symmetry between the two sets is broken.  In particular, we are going to assume $N_{1}\gg N_{2}$,  so that the extra edge is in the bigger set.  So, in conducting a search what we are trying to do is to find an extra edge in set $1$.

We need to define a quantum walk on this graph.  There are now three sets of transmission and reflection coefficients.  We have $t_{1}=2/N_{2}$ and $r_{1}=(N_{2}-2)/N_{2}$ for the vertices $\{ 3,4,\ldots N_{1}\}$,
\begin{equation}
U|j,k\rangle = -r_{1}|k,j\rangle + t_{1}\sum_{l=N_{1}+1, l\neq j}^{N_{1}+N_{2}} |k,l\rangle ,
\end{equation}
where $N_{1}+1 \leq j \leq N_{1}+N_{2}$ and $3\leq k \leq N_{1}$, and we have $t_{2}=2/N_{1}$ and $r_{2}=(N_{1}-2)/N_{1}$ for the vertices in set $2$
\begin{equation} 
U|j,k\rangle = -r_{2}|k,j\rangle + t_{2}\sum_{l=1, l\neq j}^{N_{1}} |k,l\rangle ,
\end{equation}
where $1 \leq j \leq N_{1}$ and $N_{1}+1\leq k \leq N_{1}+N_{2}$.  Finally, we have the transmission and reflection coefficients for the vertices attached to the extra edge, $\tilde{t}=2/(N_{2}+1)$ and $\tilde{r}=(N_{2}-1)/(N_{2}+1)$,
\begin{eqnarray} 
U|j,1\rangle & = & -\tilde{r}|1,j\rangle + \tilde{t}|1,2\rangle + \tilde{t}\sum_{l=N_{1}+1,l\neq j}^{N_{1}+N_{2}} 
|1,l\rangle ,  \nonumber \\
U|j,2\rangle & = & -\tilde{r}|2,j\rangle + \tilde{t}|2,1\rangle + \tilde{t}\sum_{l=N_{1}+1,l\neq j}^{N_{1}+N_{2}} 
|2,l\rangle  ,
\end{eqnarray}
where $N_{1}+1 \leq j \leq N_{1}+N_{2}$ and
\begin{eqnarray}
U|1,2\rangle & = & -\tilde{r}|2,1\rangle + \tilde{t} \sum_{l=N_{1}+1}^{N_{1}+N_{2}} |2,l\rangle \nonumber \\
U|2,1\rangle & = & -\tilde{r}|1,2\rangle + \tilde{t} \sum_{l=N_{1}+1}^{N_{1}+N_{2}} |1,l\rangle  .
\end{eqnarray}
We find that in this case there is an invariant subspace of dimension five in which the walk takes place.

Our next step is to find the characteristic equation for the resulting $5\times 5$ matrix for $U$ restricted to that subspace, and then, in order to find the zeroth order solution, take the limit as the number of vertices goes to infinity.  Now, however, we have two parameters, $N_1$ and $N_2$, so there are different ways in which we could let the number of vertices go to infinity.   As was mentioned earlier, we shall look at the case $N_{1}\rightarrow \infty$ and $N_{2}$ fixed for our zeroth order solution and then calculate corrections to it.  This result will correspond to the case $N_{1}\gg 1$.

We now start the walk in the state
\begin{equation}
|\psi_{init}\rangle  =  \frac{1}{\sqrt{2N_{1}N_{2}}} \sum_{k=1}^{N_{1}}\sum_{j=N_{1}+1}^{N_{1}+N_{2}} (|j,k\rangle
-|k,j\rangle ) ,
\end{equation}
and let it go for 
\begin{equation}
n=\frac{\pi}{4}\sqrt{N_{1}(N_{2}+2)} ,
\end{equation} steps.  We then find that the particle is on the extra edge, with a probability of $N_{2}/(N_{2}+2)$ and on one of the edges connected to the vertices linked by the extra edge with a probability of $2/(N_{2}+2)$.  As usual, we assume that when we measure the position of the particle, we do not have access to the extra edge, so that if the particle is on the extra edge, we will simply not find it.  Classically, the adjacency list for this graph contains $2N_{1}N_{2}+2$ elements, and since we know that the extra edge is in set $1$, we would only have to search half of them, i.e.\ the entries corresponding to the vertices in set $1$.  Quantum mechanically, after making approximately $\sqrt{N_{1} N_{2}}$ steps, our probability of ending up on an edge connected to one of the vertices connected to the extra edge is $2/(N_{2}+2)$, so in order to have a high probability of ending up on such an edge, we would have to repeat the walk approximately $N_{2}$ times, for a total number of steps of order $N_{2}\sqrt{N_{1}N_{2}}$.  The ratio of the total number of steps in the quantum walks to the number of items in the adjacency list is approximately $\sqrt{N_{2}/N_{1}}$. Therefore, if $N_{1}\gg N_{2}$ quantum walk gives us an advantage.  

Suppose, however, that we are faced with a different problem.  We are given a complete bipartite graph that may, or may not, have an extra edge in set $1$ that renders it no longer bipartite.  What we would like to determine is whether there is an extra edge or not, and we don't care where it is.  In that case, we only need to run the quantum walk a number of times of order one.  If after such a walk, we cannot find the particle, we know it is on the extra edge, so the graph does indeed possess such an edge.  This would require $\sqrt{N_{1}N_{2}}$ steps of a quantum walk.  Classically we would still have to search the adjacency list, which has approximately $N_{1}N_{2}$ items..  For this problem, the quantum quantum walk provides a greater speedup, the ratio of the number of steps of the quantum walk to the items in the adjacency list is $(N_{1}N_{2})^{-1/2}$, but the classical procedure will give us more information.  It will tell us where the edge is, while the quantum procedure will only tell us whether there is such an edge.

\section{Conclusion}
We have studied a number of examples in which a quantum walk can find a structural anomaly in a graph more efficiently than can a classical search.  In most cases, the anomaly was an extra edge or a set of extra edges.  The case of two stars was somewhat different in that there we found which two edges were linked, or, looking at it from the point of view of vertices, which vertex had two edges instead of one edge attached to it.

These examples suggest that there is a class of graphs whose structure can be usefully and efficiently probed by quantum walks.  What the general features of this class are, we do not know.  It would be useful to identify features of a graph that would indicate that some of its properties can be ascertained by running a quantum walk on it.  That remains a topic for future work.

Recently there has  been considerable experimental work on quantum walks on a number of different systems \cite{PeLaPoSoMoSi08}-\cite{schreiber}.  All but the last of these have been quantum walks on a line.  The paper by Schreiber, \emph{ et al.} reported on an implementation of a two-dimensional quantum walk \cite{schreiber}.  The rapid progress in this area leads us to hope that walks on more complicated geometries can be implemented, which would open the door to performing quantum walk searches.  This could make some of the results presented in this paper accessible to experiment.

\section*{Acknowledgments}
This work was supported by the National Science Foundation under grant PHY-0903660.  D.~R.\ acknowledges 
support from the European project COQUIT.

\section*{Appendix}
\subsection{One missing loop}
As was noted in the main body of the paper, this walk takes place in a five-dimensional invariant subspace spanned by the orthonormal basis $\{ |0,1\rangle , |1,0\rangle , |\psi_{L}\rangle , |\psi_{1}\rangle , |\psi_{2}\rangle \}$.  In this basis, ordered as in the previous sentence, we have that 
\begin{equation}
U_{S} =\left( \begin{array}{ccccc} 0 & -r & 0 & t\sqrt{N-1} & 0 \\ e^{i\phi} & 0 & 0 & 0 & 0 \\ 0 & t\sqrt{N-1} & 0 & r & 0 \\
0 & 0 & 0 & 0 & 1 \\ 0 & 0 & 1& 0 & 0 \end{array} \right) .
\end{equation}
The characteristic equation of $U_{S}$ is
\begin{equation}
\lambda^{5} + re^{i\phi}\lambda^{3} + -r\lambda^{2} - e^{i\phi} = 0
\end{equation}
which in the $N\rightarrow \infty$ limit becomes $(\lambda^{3} -1)(\lambda^{2}+e^{i\phi})=0$.  From this we see that $1$ will be a double root of this equation if $\phi =\pi$, $e^{2\pi i/3}$ will be a double root if $\phi = \pi /3$, and $e^{-2\pi i/3}$ becomes a double root if $\phi = -\pi /3$.  For the case $\phi = \pi$ we find that the relevant eigenvalues and eigenstates are
\begin{eqnarray}
|v_{+}\rangle = \frac{1}{2}\left( \begin{array}{c} 1 \\ -1 \\ i\sqrt{2/3} \\ i\sqrt{2/3} \\ i\sqrt{2/3} \end{array}\right) & 
{\rm for}\  \lambda = 1+i \sqrt{\frac{t}{3}}  \nonumber \\
|v_{-}\rangle = \frac{1}{2}\left( \begin{array}{c} 1 \\ -1 \\ -i\sqrt{2/3} \\ -i\sqrt{2/3} \\ -i\sqrt{2/3} \end{array}\right) & 
{\rm for}\  \lambda = 1-i \sqrt{\frac{t}{3}} .
\end{eqnarray}
We now choose 
\begin{eqnarray}
|\psi_{init}\rangle & = & \frac{1}{\sqrt{3N}} \sum_{j=0}^{N}(|0,j\rangle + |j,0\rangle + |l_{j}\rangle ) \nonumber \\
 & = & \frac{-i}{\sqrt{2}} (|v_{+}\rangle - |v_{-}\rangle ) + O(N^{-1/2}) ,
\end{eqnarray}
and, setting $\theta = \sqrt{t/3}$, this yields
\begin{equation}
U^{n}|\psi_{init}\rangle = \frac{1}{\sqrt{2}}\left( \begin{array}{c} \sin (n\theta ) \\ -\sin (n\theta ) \\  \sqrt{2/3} 
\cos (n\theta ) \\ \sqrt{2/3}\cos (n\theta ) \\ \sqrt{2/3} \cos (n\theta )  \end{array} \right) + O(N^{-1/2}) .
\end{equation}
From this equation we see that when $n\theta = \pi /2$, the particle is, with probability close to one, located on the edge with the dummy loop.  The cases $\phi = \pm \pi /3$ yield different eigenvalues and eigenstates, and hence require different initial states, but the results are qualitatively the same, the particle is with a probability close to one on the edge connected to the dummy loop after $O(N^{1/2})$ steps.

\subsection{Star graph with a clique}
This problem possesses a five-dimensional invariant subspace, $S$.  Define
\begin{eqnarray}
|\psi_{1}\rangle & = & \frac{1}{\sqrt{M}} \sum_{j=1}^{M} |0,j\rangle \nonumber \\
|\psi_{2}\rangle & = &  \frac{1}{\sqrt{M}} \sum_{j=1}^{M} |j,0\rangle \nonumber \\
|\psi_{3}\rangle & = & \frac{1}{\sqrt{M(M-1)}} \sum_{j=1}^{M}\sum_{k=1,k\neq j}^{M} |j,k\rangle \nonumber \\
|\psi_{4}\rangle & = & \frac{1}{\sqrt{N-M}} \sum_{j=M+1}^{N} |0,j\rangle \nonumber \\
|\psi_{5}\rangle & = & \frac{1}{\sqrt{N-M}} \sum_{j=M+1}^{N} |j,0\rangle .
\end{eqnarray}
These vectors are orthonormal and constitute a basis for $S$.  With this ordering, the matrix for $U_{S}$ is
\begin{equation}
\left( \begin{array}{ccccc} 0 & tM-1 & 0 & 0 & t\sqrt{M(N-M)} \\ -\tilde{r} & 0 & \tilde{t}\sqrt{M-1} & 0 & 0 \\
\tilde{t}\sqrt{M-1} & 0 & \tilde{r} & 0 & 0 \\ 0 & t\sqrt{M(N-M)} & 0 & 0 & 1-tM  \\ 0 & 0 & 0 & 1 & 0 \end{array} \right) .
\end{equation}
The characteristic polynomial for this matrix is 
\begin{eqnarray}
\lambda^{5}+(\tilde{t}-1)\lambda^{4} + [2(M-1)t + \tilde{t}-2]\lambda^{3} \nonumber \\
 - [2(M-1)t + \tilde{t}-2]\lambda^{2} -(\tilde{t}-1)\lambda -1 =0 ,
\end{eqnarray}
which in the $N\rightarrow \infty$ limit becomes
\begin{equation}
\lambda^{5} + (\tilde{t}-1)\lambda^{4} + (\tilde{t}-2)\lambda^{3}-(\tilde{t}-2)\lambda^{2} - (\tilde{t}-1)\lambda -1=0 .
\end{equation}
The $N\rightarrow \infty$ equation has a double root of $-1$.  Setting $\lambda = -1+\delta \lambda$ we find that
\begin{equation}
\delta\lambda = \pm i \sqrt{\frac{2M(M-1)}{(2M-1)N}}  \equiv \pm i\theta .
\end{equation}
The eigenvectors corresponding to these eigenvalues are 
\begin{equation}
|v_{+}\rangle = \sqrt{\frac{M-1}{2(2M-1)}} \left( \begin{array}{c} 1 \\ 1 \\ -1/\sqrt{M-1} \\ -i\sqrt{(2M-1)/(2M-2)} \\
i\sqrt{(2M-1)/(2M-2)} \end{array} \right) ,
\end{equation}
for $ \lambda = -1+i\theta$, and 
\begin{equation} 
|v_{-}\rangle = \sqrt{\frac{M-1}{2(2M-1)}} \left( \begin{array}{c} 1 \\ 1 \\ -1/\sqrt{M-1} \\ i\sqrt{(2M-1)/(2M-2)} \\
-i\sqrt{(2M-1)/(2M-2)} \end{array} \right) 
\end{equation}
for $\lambda = -1-i\theta$.
For the initial state we choose 
\begin{eqnarray}
|\psi_{init}\rangle & = & \frac{1}{\sqrt{2N}} \sum_{j=1}^{N} (|0,j\rangle - |j,0\rangle ) \nonumber \\
 & = & \frac{i}{\sqrt{2}}(|v_{+}\rangle - |v_{-}\rangle ) +O( \sqrt{M/N}) .
\end{eqnarray}
We then find that up to terms of order $(M/N)^{1/2}$, 
\begin{equation}
U^{n}|\psi_{init}\rangle = (-1)^{n}\sqrt{\frac{M-1}{2M-1}}\left( \begin{array}{c} \sin (n\theta ) \\ \sin (n\theta ) \\
-(M-1)^{-1/2}\sin (n\theta ) \\ \sqrt{\frac{2M-1}{2(M-1)}} \cos (n\theta ) \\ -\sqrt{\frac{2M-1}{2(M-1)}} \cos (n\theta )
\end{array} \right) .
\end{equation}
Therefore, when $n\theta = \pi /2$ we find that the particle is on one of the edges going from the central vertex to the clique with a probability of $(2M-2)/(2M-1)$ and a probability of being on the clique itself of $1/(2M-1)$.

\subsection{Two stars}
The dimension of this problem can be reduced still further, from $8$ to $4$.  If we define the vectors
\begin{eqnarray}
|w_{1}\rangle & = & \frac{1}{\sqrt{2}} (|\psi_{1}\rangle - |\psi_{3}\rangle ) \nonumber \\
|w_{2}\rangle & = & \frac{1}{\sqrt{2}} (|\psi_{5}\rangle - |\psi_{7}\rangle ) \nonumber \\
|w_{3}\rangle & = & \frac{1}{\sqrt{2}} (|\psi_{2}\rangle - |\psi_{4}\rangle ) \nonumber \\
|w_{4}\rangle & = & \frac{1}{\sqrt{2}} (|\psi_{6}\rangle - |\psi_{8}\rangle )  
\end{eqnarray}
then we find that the action of the unitary operator, U, is given by
\begin{eqnarray}
U|w_{1}\rangle & = & -|w_{3}\rangle \nonumber \\
U|w_{2}\rangle & = & |w_{4}\rangle \nonumber \\
U|w_{3}\rangle & = & -r|w_{1}\rangle +t\sqrt{N-1}|w_{2}\rangle \nonumber \\
U|w_{4}\rangle & = & r|w_{2}\rangle + t\sqrt{N-1} |w_{1}\rangle .
\end{eqnarray}
Therefore, if our initial state is in the subspace, $S^{\prime}$, which we define to be the linear span of the vectors $\{ |w_{j}\rangle | j=1,\dots 4 \}$, then the dynamics can be describe completely within this four-dimensional subspace.  The initial state  
\begin{eqnarray}
|\psi_{init}\rangle & = & \frac{1}{2\sqrt{N}} \left[ \sum_{j=1}^{N} ((A,j\rangle + |j,A\rangle ) - (|1,B\rangle 
+ |B,1\rangle ) \right.  \nonumber \\
& & \left. - \sum_{j=N+1}^{2N-1}(|B,j\rangle + |j,B\rangle ) \right]  ,
\end{eqnarray}
can be expressed as
\begin{eqnarray}
|\psi_{init}\rangle & = & \frac{1}{\sqrt{2N}}(|w_{1}\rangle + |w_{3}\rangle ) \nonumber \\
 & & + \sqrt{\frac{N-1}{2N}} (|w_{2}\rangle + |w_{4}\rangle ) ,
 \end{eqnarray}
 so that it is, in fact, in $S^{\prime}$.  Therefore, we have reduced our problem to a four-dimensional one.
 
We can go even further if we look at $U^{2}$.  Under the action of $U^{2}$, the subspace $S^{\prime}$ splits into two two-dimensional subspaces, one spanned by $\{ |w_{1}\rangle , |w_{2}\rangle \}$ and the other spanned by $\{ |w_{3}\rangle , |w_{4}\rangle \}$.  In particular, we have that
 \begin{eqnarray}
 U^{2}|w_{1}\rangle & = & r|w_{1}\rangle - t\sqrt{N-1} |w_{2}\rangle \nonumber \\
 U^{2}|w_{2}\rangle & = & r|w_{2}\rangle + t\sqrt{N-1} |w_{1}\rangle ,
 \end{eqnarray}
which means that in the $\{ |w_{1}\rangle , |w_{2}\rangle \}$ subspace $U^{2}$ can be described by the matrix
\begin{equation}
U^{2} = \left( \begin{array}{cc} r & t\sqrt{N-1} \\ -t\sqrt{N-1} & r \end{array} \right) .
\end{equation}
Similarly, we have that
 \begin{eqnarray}
 U^{2}|w_{3}\rangle & = & r|w_{3}\rangle + t\sqrt{N-1} |w_{4}\rangle \nonumber \\
 U^{2}|w_{4}\rangle & = & r|w_{4}\rangle - t\sqrt{N-1} |w_{3}\rangle ,
 \end{eqnarray}
which means that in the $\{ |w_{3}\rangle , |w_{4}\rangle \}$ subspace $U^{2}$ can be described by the matrix
\begin{equation}
U^{2} = \left( \begin{array}{cc} r & -t\sqrt{N-1} \\ t\sqrt{N-1} & r \end{array} \right) .
\end{equation}
So what we are left with are two two-dimensional problems,both of which are, mathematically, equivalent to Grover searches.  The eigenvalues of both matrices are $\lambda = r \pm it\sqrt{N-1}$, which we shall denote by $e^{\pm i\theta}$, respectively.  This implies that $\theta \simeq 2/\sqrt{N}$.  It is straightforward to find the eigenvectors and to use them to raise the matrices to an arbitrary power.  We find that
\begin{eqnarray}
U^{2n}|w_{1}\rangle & = & \cos (n\theta )|w_{1}\rangle -\sin (n\theta )|w_{2}\rangle \nonumber \\
U^{2n}|w_{2}\rangle & = & \sin (n\theta )|w_{1}\rangle +\cos (n\theta )|w_{2}\rangle \nonumber \\
U^{2n}|w_{3}\rangle & = & \cos (n\theta )|w_{3}\rangle +\sin (n\theta )|w_{4}\rangle \nonumber \\
U^{2n}|w_{4}\rangle & = & -\sin (n\theta )|w_{3}\rangle +\cos (n\theta )|w_{4}\rangle .
\end{eqnarray}
These equations imply that when $n\theta =\pi /2$ our initial state, $|\psi_{init}\rangle$, will have been transformed into $(1/\sqrt{2})(|w_{1}\rangle - |w_{3}\rangle )$ up to terms of order $N^{-1/2}$.  That means the probability that the particle is located on the edges where the stars are connected is almost one.  The condition $n\theta =\pi /2$ implies that $n=\pi\sqrt{N}/4$, and the number of steps in the walk is twice that, or $\pi\sqrt{N}/2$.

Note that other initial states for this walk are possible.  The state
\begin{equation}
|\psi_{init}^{\prime}\rangle = \frac{1}{\sqrt{N}}|w_{1}\rangle + \sqrt{\frac{N-1}{N}}|w_{2}\rangle ,
\end{equation}
which is an equal superposition of all of the outgoing states on the first star minus all the outgoing states on the second, will also lead to a successful search.  After $O(\sqrt{N}$ steps the particle will end up, to very good approximation, in outgoing states on the connected edges. 

\subsection{Complete bipartite graph}
Define the orthonormal set
\begin{eqnarray}
|\psi_{1}\rangle & = & \frac{1}{\sqrt{2}}(|1,2\rangle + |2,1\rangle ) \nonumber \\
|\psi_{2}\rangle & = & \frac{1}{\sqrt{2N_{2}}} \sum_{j=N_{1}+1}^{N_{1}+N_{2}}(|j,1\rangle + |j,2\rangle ) \nonumber \\
|\psi_{3}\rangle & = & \frac{1}{\sqrt{2N_{2}}} \sum_{j=N_{1}+1}^{N_{1}+N_{2}}(|1,j\rangle + |2,j\rangle ) \nonumber \\
|\psi_{4}\rangle & = & \frac{1}{\sqrt{(N_{1}-2)N_{2}}}  \sum_{k=3}^{N_{1} }\sum_{j=N_{1}+1}^{N_{1}+N_{2}}
|j,k\rangle \nonumber \\
|\psi_{5}\rangle & = & \frac{1}{\sqrt{(N_{1}-2)N_{2}}}  \sum_{k=3}^{N_{1}} \sum_{j=N_{1}+1}^{N_{1}+N_{2}}
|k,j\rangle .
\end{eqnarray}
They span an invariant subspace, $S$, of $U$, the operator that advances the walk one step on a complete bipartite graph with an extra edge.  The matrix of the restriction of $U$ to $S$, $U_{S}$ is given by
\begin{equation}
U_{S}=\left(\begin{array}{ccccc} -\tilde{r} & \tilde{t}\sqrt{N_{2}} & 0 & 0 & 0 \\ 0 & 0 & -(r_{2}-t_{2}) & 0 & 
2\sqrt{t_{2}r_{2}} \\ \tilde{t}\sqrt{N_{2}} & \tilde{r} & 0 & 0 & 0 \\ 0 & 0 & 2\sqrt{t_{2}r_{2}} & 0 & r_{2}-t_{2} \\
0 & 0 & 0 & 1 & 0 \end{array} \right) .
\end{equation}
The characteristic equation of this matrix is
\begin{equation}
(\lambda -1)\{ \lambda^{4} + (\tilde{r}+1)\lambda^{3} +[\tilde{r}+1 -\tilde{t}(r_{2}-t_{2})]\lambda^{2} +(\tilde{r}+1)
\lambda +1 \} =0 .
\end{equation}
This problem has two parameters, $N_{1}$ and $N_{2}$, so there are different  ways to take the limit as the number of vertices goes to infinity.  We shall consider the case $N_{1}\rightarrow \infty$ and $N_{2}$ fixed.  This implies that to obtain our zeroth order solution we will let $t_{2}\rightarrow 0$ and $r_{2}\rightarrow 1$.  In this limit, the fourth order equation for $\lambda$ becomes
\begin{equation}
\lambda^{4} + (\tilde{r}+1)\lambda^{3} + 2\tilde{r}\lambda^{2} + (\tilde{r}+1)\lambda + 1=0 .
\end{equation}
We find that $-1$ is a double root of this equation, so we set $\lambda = -1 +\delta\lambda$ and substitute it into the actual characteristic equation keeping only the smallest terms.  This gives us $\delta\lambda =\pm i\theta$ where
\begin{equation}
 \theta = \sqrt{\frac{2t_{2}}{N_{2}+2}} .
\end{equation}
The corresponding eigenvectors are 
\begin{equation}
|v_{+}\rangle = \frac{1}{\sqrt{2(N_{2}+2)}} \left(\begin{array}{c} -\sqrt{N_{2}} \\ 1 \\ 1 \\ -i\sqrt{(N_{2}+2)/2} \\ 
i\sqrt{(N_{2}+2)/2} \end{array}\right) ,
\end{equation}
for $\lambda = -1 +i\theta$, and
\begin{equation}
|v_{-}\rangle = \frac{1}{\sqrt{2(N_{2}+2)}} \left(\begin{array}{c} -\sqrt{N_{2}} \\ 1 \\ 1 \\ i\sqrt{(N_{2}+2)/2} \\ 
-i\sqrt{(N_{2}+2)/2} \end{array}\right) ,
\end{equation}
for $\lambda = -1 -i\theta$.  Both of these expressions are valid up to corrections of order $N_{1}^{-1/2}$.

For our initial state we choose 
\begin{eqnarray}
|\psi_{init}\rangle & = & \frac{1}{\sqrt{2N_{1}N_{2}}} \sum_{k=1}^{N_{1}}\sum_{j=N_{1}+1}^{N_{1}+N_{2}} (|j,k\rangle
-|k,j\rangle ) \nonumber \\
 & = & \frac{i}{\sqrt{2}}(|v_{+}\rangle - |v_{-}\rangle ) + O(N_{1}^{-1/2}) . 
\end{eqnarray}
After $n$ steps, the state of the system is
\begin{equation}
U^{n}|\psi_{init}\rangle = \left( \begin{array}{c} -\sqrt{\frac{N_{2}}{N_{2}+2}} \sin (n\theta ) \\ \frac{1}{\sqrt{N_{2}+2}}
\sin (n\theta ) \\ \frac{1}{\sqrt{N_{2}+2}} \sin (n\theta ) \\ \frac{1}{\sqrt{2}} \cos (n\theta ) \\ -\frac{1}{\sqrt{2}} \cos 
(n\theta ) \end{array} \right) .
\end{equation}
Note that when $n\theta = \pi /2$ the particle is on the extra edge, with a probability of $N_{2}/(N_{2}+2)$ and
on one of the edges connected to the vertices linked by the extra edge with a probability of $2/(N_{2}+2)$.

\end{document}